\documentclass[review,sort&compress,11pt]{elsarticle}

\usepackage{lineno,hyperref,soul,multirow}
\usepackage[utf8]{inputenc}
\usepackage{amsmath,amssymb,color,xcolor,gensymb}
\usepackage{graphicx,caption,subcaption} 
\usepackage{amsfonts,amsthm,amsthm,MnSymbol} 
\usepackage{graphicx,float} 
\modulolinenumbers[5]

\usepackage{xcolor}
\usepackage{geometry}
\begin{document}
\begin{frontmatter}

\title{Tailoring negative pressure by crystal defects: Crack induced hydride formation in Al alloys}

\author{A. Tehranchi,$^{1\ast}$} 
\author{P. Chakraborty,$^{1}$}
\author{M.  L\'opez Freixes$^{1}$}
\author{ E. McEniry$^{1}$}
\author{B. Gault$^{1,2}$} 
\author{T. Hickel $^{1}$} 
\author{J. Neugebauer $^{1}$} 
\address{1-Max-Planck-Institut f\"ur Eisenforschung GmbH, D\"usseldorf,D-40237, Germany}
\address{2-Department of Materials, Imperial College, South Kensington,London, SW7 2AZ, United Kingdom}

\begin{abstract}
  Climate change motivates the search for non-carbon-emitting energy generation and storage solutions. Metal hydrides show promising characteristics for this purpose. They can be further stabilized by tailoring the negative pressure of microstructural and structural defects. Using systematic ab initio and atomistic simulations, we demonstrate that an enhancement in the formation of hydrides at the negatively pressurized crack tip region is feasible by increasing the mechanical tensile load on the specimen. The theoretical predictions have been used to reassess and interpret atom probe tomography experiments for a high-strength 7XXX-aluminium alloy that show a substantial enhancement of hydrogen concentration at structural defects near a stress-corrosion crack tip. These results contain important implications for enhancing the capability of metals as H-storage materials.
\end{abstract}

\end{frontmatter}

%\linenumbers
\section{Introduction}
While fossil fuels have been a pillar of the rapid growth of modern human society, the ever growing consumption of coal, crude oil, natural gas, and their derivatives inflicts several challenges upon the human civilization~\cite{smil2018energy}. Since the overall rise of global temperature~\cite{olivier2012trends} and climate changes are inevitable if the current rate of fossil fuel consumption is maintained, diversification of fuels is one of the most prominent responses to this crisis. 
Using hydrogen as a non-carbon-emitting energy carrier is a very promising way of energy storage and carriage~\cite{dawood2020hydrogen}. Metal hydrides (MH$_x$) are believed to be among the most technologically relevant classes of hydrogen storage materials, because of their capability to be used in a diverse set of applications such as electrochemical cycling, thermal storage, heat pumps, and purification/separation~\cite{klebanoff2012final}. The accumulation of H atoms is usually associated with an increase in the specific volume of the hydride phase.  Thus, it can be envisioned that the presence of a \emph{negative pressure} field will enhance the formation of these hydrides.  Moreover, in contrast to ample studies on metal hydrides in the high (positive) pressure domain of phase diagrams~\cite{smith2021isotherm}, the negative pressure domain of phase diagrams remains largely unexplored due to experimental limitations  (c.f.~\cite{imre2007existence,mcmillan2003new,mcmillan2002new,siol2018negative}). 

A possible way to overcome the experimental challenges is to carefully use crystal defects as laboratory tools to induce negative pressure at small scales. At this scale, the material can undergo stresses up to few GPa without loss of integrity.  This negative pressure is present in close vicinity of edge dislocation, or precipitates with larger atomic volume such as Al$_3$Sc precipitates ~\cite{gupta2021combined} in pure Al.  However, in both cases the magnitude of negative pressure cannot be adjusted externally. 

Among the crystal defects, the stress field of cracks is not only a function of the material properties, but also of the external loading. This is different to the stress fields of dislocations, precipitates and grain boundaries, which are fully determined by the material and geometrical properties, but can hardly be changed during service. Thus one can tailor the magnitude of the negative pressure in the close neighborhood of the crack. On the one hand, it can thus be envisioned that the negative pressure domain of the phase diagrams can be explored by exploiting cracks. On the other hand, this mechanism might even give the chance to control the conditions of hydride formation during charging and hydrogen release during discharging.

In this work, we demonstrate with ab initio and empirical potential based simulations the theoretical feasibility of using cracks as laboratory tools for inducing negative pressure and enhancing hydride formation. Based on the achieved theoretical insights, we reassess novel experimental evidence of the formation of high H concentration regions at crack tips using atom probe tomography (APT) and validated the theoretical predictions.  

The study is performed for aluminum, since it is a light (with $Z=13$), cheap and abundant element that seems to be a promising metal hydride former.
Among the hydride phases in Al, the AlH$_3$, Alane hydride, is very efficient having $\approx 11$\% H gravimetric density,  making this hydride ``near perfect" for H-storage applications~\cite{jiang2021alh3}. Al is not an intrinsically brittle material as compared to the other light hydride-forming candidates such as Mg, and Li, making it even more attractive for hydrogen storage solutions. However, the hydrogenation of defect-free Al and the formation of metal hydrides at finite temperatures requires extremely high hydrogen gas pressures, which is one of the major challenges for the application of this hydride~\cite{graetz2013recent,graetz2011aluminum,jiang2021alh3}.
In this work we suggest to meet this challenge by exploiting the concept of defect induced negative pressure.

\section{Thermodynamics of hydride formation in Al}\label{sec:thermo}

\subsection{Computational methodology of ab initio calculations}
Density functional theory (DFT) calculations were carried out using the projector augmented wave (PAW) potentials as implemented in the Vienna \textit{Ab initio} Simulation Package (VASP)~\cite{kresse1993ab,kresse1996efficiency,kresse1996efficient}. The exchange and correlation terms were described by the generalized gradient approximation (GGA) proposed by Perdew, Burke, and Ernzerhof (PBE)~\cite{perdew1996generalized}. A plane-wave cut-off of 500 eV was taken for all calculations. The convergence tolerance of the atomic force is 0.01 eV/{\AA}  and the total energy is $10^{-6}$ eV. Brillouin-zone integration was made using Methfessel–Paxton~\cite{Methfessel1989} smearing. Ionic relaxations were allowed in all calculations keeping the shape and volume fixed. The equilibrium structure of Al with a lattice parameter of 4.04 {\AA} obtained within the convergence criteria is consistent with previous DFT-GGA calculations and has been used to construct the supercells.

\subsection{Solubility of H in Al}

The solution energy of an H atom in the Al matrix is determined using DFT.
Since we are interested in H charging processes, which are usually expressed in terms of the H$_2$ gas pressure, we consider the chemical potential of the H$_2$ molecule $\mu_{\rm H}^0=\frac{1}{2}E(H_2)$ as a reference. We find the tetrahedral sites more favorable for H (solution energy of 0.743 eV) than octahedral sites (0.845 eV), in accordance with previous studies in the literature~\cite{wolverton2004hydrogen}.  
Further, we reveal significant first (-0.114 eV) and weaker second (-0.028 eV) nearest neighbor attractive interactions between two H atoms in the tetrahedral sites of Al (cf.\ Table~\ref{Tab:int} in the ~\ref{App:A}).  Attractive interaction energies that are larger than $k_{\rm B}T$ at room temperature ($\approx$ 0.03 eV) facilitate the population of H atoms in the defected regions.  As previously shown for Ni ~\cite{von2011hydrogen}, nanohydrides would not exist in defected regions without the existence of these interactions, even for the maximum relevant H bulk concentrations, $C_{\rm b}$, in Al ($\approx 1000$~appm)~\cite{birnbaum1997hydrogen}. 
Though attractive interactions are also determined for pairs containing octahedral sites, they cannot compensate for the energetic preference of the tetrahedral sites indicated above. 
 
Due to the extensive computational costs, the large scale atomistic simulations required for defects such as cracks are not possible using DFT. In this work, the interaction between atoms next to cracks is therefore described by an angular dependent potential (ADP)~\cite{apostol2010angular}. For the original version of the ADP the H--H interaction energies are significantly lower than those obtained via DFT (cf.\ Table~\ref{Tab:int}). To include these crucial features we augmented the ADP by the addition of an attractive Morse-type interaction between H atoms, which we call ADMP (see~\ref{App:C}). 

\subsection{Formation energies and excess volumes of different hydride phases}
Our study of hydrides in Al is focused on the most stable candidates at each stoichiometry, i.e., the rock-salt type ${\rm AlH}$, flourite type ${\rm AlH}_2$, and $\alpha$-Alane hydride ${\rm AlH}_3$~\cite{wolverton2004hydrogen}.  
To determine the influence of stress, the excess volume of the hydrides is important, i.e., the difference between the fully relaxed volume of the hydride supercell  $V({\rm Al}_m{\rm H}_n)$ and of the metal, $V(\rm Al)=16.48\text{\AA}^3$, per Al atom at zero pressure. The DFT-determined excess volume of AlH$_3$ is particularly large ($\approx$ 18 \AA$^3$) and thus its interaction with the tensile stress field of the cracks will be considerable (Table~\ref{Tab:int}). In the case of AlH$_2$, the excess volume remains positive, but its magnitude is $\approx 5$ times smaller than that of AlH$_3$. 
Interestingly, in the case of rock salt AlH, the excess volume is very small but negative. 
We conclude that the effect of the stress fields on the hydride formations is only significant for AlH$_3$. The ADMP follows the qualitative trend of the excess volumes of the hydride, however with less magnitude. We also calculated the excess volume of a single H atom in Al bulk ($\Delta V$(H) =1.808 $\text{\AA}^3$) by constructing the pressure-concentration plots and calculating the excess volume using their slope \cite{tehranchi2017softening}, since it is important to trigger long-range diffusion of H atoms in the bulk towards the defected regions. A comparison of this excess volume with the excess volume per H atoms in the hydrides indicates an increase by a factor of 3.25 in the case of AlH$_3$. For the AlH$_2$ and AlH this factor is 0.99 and -0.015, respectively. The reason of this  significant increase in excess volume of AlH$_3$ is the formation of covalent bonds in this covalently bonded trihydride rather than metallic bonds, which are formed in AlH and AlH$_2$~\cite{jiang2021alh3}.
This observation also demonstrates the strong effect of a negative stress field on the formation of AlH$_3$ out of H populated regions.

\begin{figure}
\centering
     \includegraphics[width=1.0\textwidth]{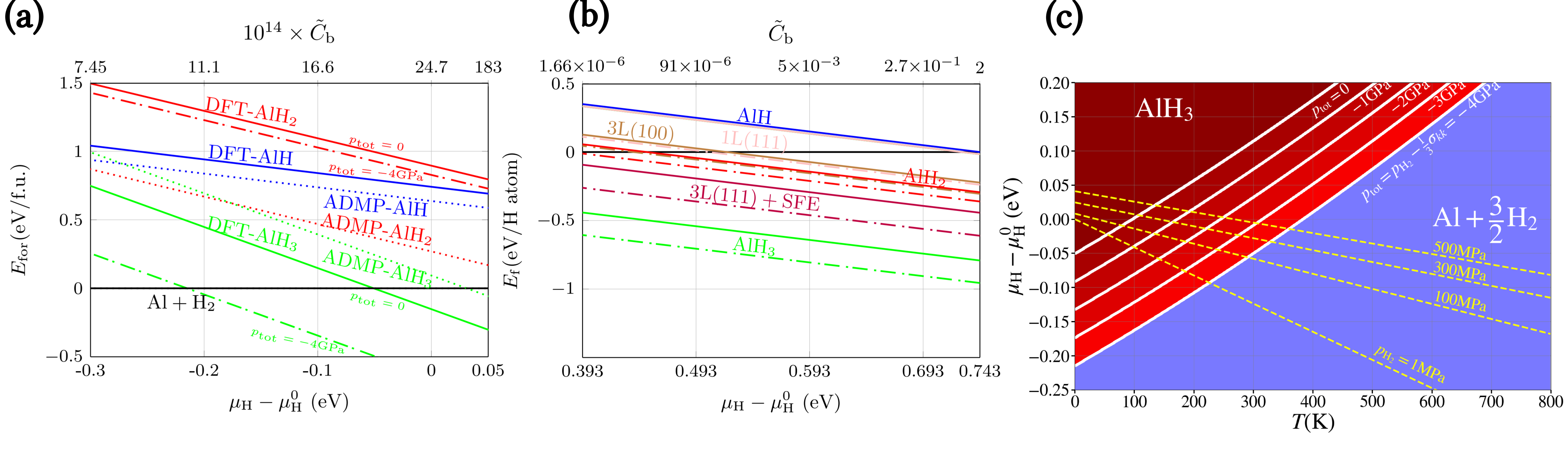}
    \caption{(a) Formation energy per H atom of the different Al hydride phases versus the hydrogen chemical potential based on DFT and ADMP simulations.
    The solid, dashed, and dotted lines denote results for DFT at zero total pressure, DFT at negative total pressure, and ADMP at zero total pressure, respectively. Green, red, and blue lines denote AlH$_3$, AlH$_2$, and AlH, respectively. The upper horizontal axis denotes, $\tilde{C}_{\rm b}$, the atomic bulk concentration of H atoms at room temperature.
    (b) Formation energy per hydrogen atom of the different Al planar and bulk hydrides phases and versus the hydrogen chemical potential based on DFT. 
    The solid, and dashed, lines denote results for DFT at zero pressure and negative pressure, respectively. 
    (c) Pressure dependent phase diagram of AlH$_3$ hydride in $\mu$--$T$ space.  The dashed lines denote the variation of the chemical potential of H$_2$ gas for various $p_{{\rm H}_2}$ obtained using Eqs.~\eqref{Eq:FH2}-\eqref{Eq:mup}.
    }
    \label{fig:phase_for}
\end{figure}

The DFT calculations for the formation energies of the hydrides per unit formula yield the $\alpha$-Alane hydride as the most stable phase at $\mu^0_{\rm H}$ (Figure~\ref{fig:phase_for}a, Table~\ref{Tab:int} ). This is also true for the empirical potentials, while $E_{\rm f}({\rm Al}{\rm H}_3)$ is closer to the DFT value after the modification of the ADP by the Morse potential. 
The remaining discrepancies, e.g., the over-stabilization of the AlH$_2$ hydride do not affect the  trends for the hydride formation.  The reason is that the DFT calculations depicted in Figure~\ref{fig:phase_for}b show that the AlH$_2$ hydride is also stable with respect to individual H atoms in the chemical potential domain in which the MD--MC simulation are performed and can serve as a nucleus for AlH$_3$.  In this regard, the simulation using ADMP provides a lower bound for the enhancement effect of the crack tip regions.

Beyond the values at $\mu_{\rm H} = \mu_{\rm H}^0$, Figure~\ref{fig:phase_for}a depicts the formation energy of different Al hydrides versus a variation of the chemical potential $\mu_{\rm H}$ for both pressure free and negative pressure cases. 
The effect of pressure $p$ is included in the model by adding the $-p\Delta V$ term to the formation energies, using the excess volume of H for $\Delta V$. Throughout this work we refer to the mechanical pressure induced by defects as $p^{(m)}$, i.e., the total pressure acting on the hydride is $p_{\rm tot}=p_{{\rm H}_2}+p^{\rm (m)}$.
It can be seen that the $\alpha$-Alane hydride remains to be the most stable hydride for all relevant values of the chemical potential, in accordance with~\cite{wolverton2004hydrogen}. 
A negative pressure field of magnitude $p_{\rm tot}=-4$ GPa significantly reduces the formation energy of $\alpha$-Alane, whereas this effect is negligible for the other hydrides. The choice $p_{\rm tot}=-4$ GPa corresponds to the maximum pressure achievable in the close (a few {\AA}ngstroms) neighbourhood of the crack tip before the emission of dislocation or decohesion~\cite{yamakov2006molecular}. Moreover, this pressure is well below the theoretical tensile strength of Al of $\sigma_{\rm max}=12$ GPa~\cite{van2004thermodynamics}.
The results of the ADMP in Figure~\ref{fig:phase_for}a also yield AlH$_3$ as the most stable phase above $\Delta \mu = - 0.175$, but the AlH$_2$ phase is more stable than the AlH phase. Since we use this potential for the simulated accumulation of individual H atoms at the crack tip region and the most stable hydride is conserved by the ADMP, this feature is not affecting the results of our study. The DFT calculations depicted in Figure~\ref{fig:phase_for}b also show that the AlH$_2$ hydride is more than lattice gas in the chemical potential regions in which the ADMP simulations will be performed,

Motivated by our experimental results, which indicate the presence of hydrides along (111) planes at the crack tip region, we computed the formation energy of several planar hydrides in fcc Al.  Figure~\ref{fig:phase_for}b depicts the formation energies per H atom of the most stable planar hydrides in addition to the bulk hydrides with respect to  the chemical potential of a single H atom in the isolated tetrahedral site in fcc Al, i.e. $\mu_{\rm H}=\mu_{\rm H}^{\rm T}=0.743~\rm eV$. 
It can be seen that the planar hydride consisting of 3 layers of H atoms  along a (111) plane containing a stacking fault is the most stable structure in comparison to all hydrides except AlH$_3$.  
The application of negative pressure does not change this observation.  
The reason for the preference of the (111) plane is that H atoms along these planes can distort the Al matrix and find the energetically favorable structure, while these distortions are absent in AlH$_2$ and planar hydrides along (100) planes.

After the consistency of the methods at $T$=0 has been confirmed, we next construct the phase diagram of the hydrides for different temperatures and chemical potentials. To this end, it is necessary to calculate the  vibrational entropy contribution to the formation free energy of the hydride phases,
\begin{align}
  F_{\rm for}({\rm Al}_m{\rm H}_n)=E_{\rm f}({\rm Al}_m{\rm H}_n)+F^{\rm vib}({\rm Al}_m{\rm H}_n)-m F^{\rm vib}({\rm Al})-\frac{n}{2}F^{\rm vib}({\rm H}_2)\label{Eq:Ffor}
\end{align}
Here, the vibrational free energy $F^{\rm vib}$ is calculated via the quasi-harmonic approximation~\cite{grabowski2007ab} (see  see~\ref{App:E}). 
The temperature dependent part of the formation energy of the H$_2$ molecule is determined using the experimental relation~\cite{wolverton2004hydrogen}
\begin{align}
  F({\rm H}_2)=\frac{7}{2}k_{\rm B}T-TS^{\rm expt}({\rm H}_2), \label{Eq:FH2}
\end{align}
where $k_{\rm B}$ is the Boltzmann constant, $T$ is the temperature, and $S^{\rm expt}({\rm H}_2)\approx 15.7k_{\rm B}$. The enthalpy and entropy of the H$_2$ molecule are evaluated at standard conditions, respectively~\cite{atkins1998physical}. 

Figure~\ref{fig:phase_for}c is the pressure-dependent phase diagram of the hydrides in terms of different temperatures and hydrogen chemical potentials. Consistent with the $T$=0 results, the $\alpha$-Alane hydride is stable in the high $\mu_{\rm H}$ regions and the competing phase is pure Al accompanied by H$_2$.  The white solid lines denote the border of the aforementioned two phases at various $p_{\rm tot}$ values. It can be seen that negative pressure significantly enhances the formation of AlH$_3$ for given temperatures as well as for given chemical potentials and no hydride AlH$_n$ with $n<3$ becomes stable.

\begin{figure}[H]\centering
 \centering
     \includegraphics[width=1.0\textwidth]{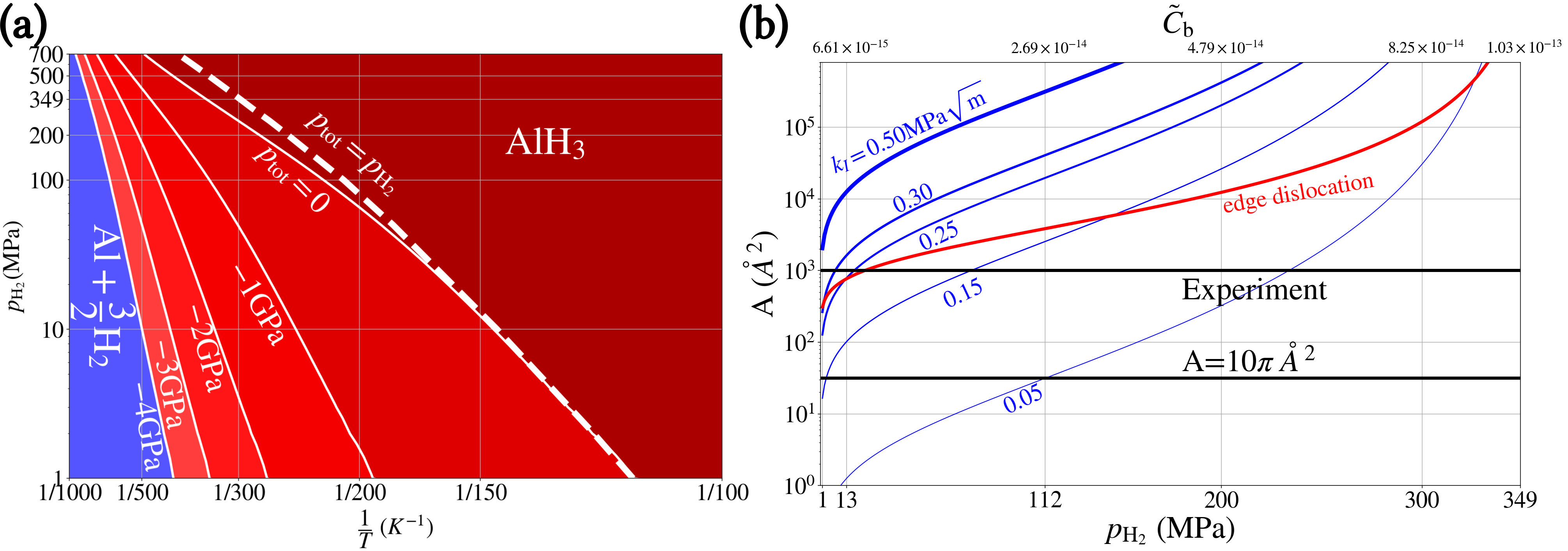}
    \caption{(a)  Pressure dependent phase diagram of Alane AlH$_3$ in the $p_{{\rm H}_2}$--$T$ space as a function of inverse temperature and external hydrogen pressure (left axis). The plot provides the critical negative mechanical pressure  for the formation of the Alane hydride, i.e. $p^{\rm (m)}=p_{\rm tot}-p_{{\rm H}_2}$. The dashed line denote the case in which no mechanical load is applied.
  (b) Dependence of the cross sectional area, $A$,  around the crack tip for which the conditions allow for hydride formation at T=300 K versus the H$_2$ pressure for different values of the cracks stress intensity factor $k_{I}$. The values of $k_{I}=0.3$ MPa$\sqrt m$ and $k_{I}=0.5$ MPa$\sqrt m$ are typical  for the emission of dislocations from the crack on a (111) plane and cleavage of the crack on non-emitting crack geometries~\cite{yamakov2014investigation}. } \label{fig:effic}
\end{figure}

The chemical potential of hydrogen can be expressed in terms of the external H$_2$ gas pressure, $p_{{\rm H}_2}$ as
\begin{align}
  \mu_{\rm H}=\mu_{\rm H}^0+\frac{1}{2}\left((p_{{\rm H}_2}-p_0)b+k_{\rm B}T\ln{\frac{p_{{\rm H}_2}}{p_0}}\right)\label{Eq:mup}
\end{align}
where $p_0=0.1$ MPa is the standard pressure \cite{san2007permeability}. The co-volume constant $b=26.3\text{\AA}^3$ accounts for the volume of a H$_2$ molecule in the Nobel-Abel equation of state. Figure~\ref{fig:effic}a illustrates the mechanical pressure dependent phase diagram of the $\alpha$-Alane hydride in $p_{{\rm H}_2}$--$T$ space. It can be seen that the presence of a negative mechanical pressure field in the Al matrix significantly reduces the required $p_{{\rm H}_2}$ for the formation of the $\alpha$-Alane hydride. This feature is the most important message of this work.

Although cracks are usually considered as detrimental defects in material science and engineering, they can be used as laboratory tool to induce negative pressure therewith supporting the formation of hydrides without causing complete failure of the material. The stress field of the crack tip is a function of the stress intensity factor, which itself is proportional to the loading and depends on the geometry of the crack. Thus, it can be envisioned that the loading on the crack provides a continuous variable to probe the negative pressure domain.  The interaction of the H excess volume and the long-range stress field of the crack guides the H atoms toward the crack tip regions and induces higher $C_{\rm H}$ regions near the crack tip, which can serve as hydride nuclei. When the nucleus is formed, it can transform to the most stable hydride. 

Cracks in Al based alloys are particularly effective in this regard, because
\begin{enumerate}
\item  
The ductile nature of pure Al combined with alloying, which usually does not show brittle fracture~\cite{yamakov2014investigation},
can reduce the damaging effects of hydrides ~\cite{ciaraldi1980studies,scully2012hydrogen} by tailoring the interface energies~\cite{zhang2021first}. Alloying and doping are,  e.g.,  used for enhancement of  nickel metal hydride batteries~\cite{tsukahara1999improvement,tsukahara1997vanadium}

\item The interaction between the excess volume of $\alpha$-Alane hydride resulting in a negative strain in the vicinity of the crack and the negative pressure at the crack tip region stemming from the tensile strain reduces the formation free energy of this hydride. 

\end{enumerate}
Therefore, the excess volume of both, individual H atoms and the $\alpha$-Alane hydride, determines the kinetics and thermodynamics of this process. If the excess volume of individual H atoms is low, then the diffusion toward the crack  will be slow. Moreover, if the excess volume of the desired hydride is low, then the decrease in its formation free energy associated with the presence of the crack stress field and subsequently the decrease in the required H$_2$ gas pressure will not be significant.

It is evident from Figure~\ref{fig:effic}a that the external H$_2$ pressure needed for the formation of the hydride at ambient conditions ($p_{\rm tot}=P_{{\rm H}_2}$ and room temperature) is $\approx 349$ MPa, which is larger than the typical economically favorable values of 35-70 MPa~\cite{wikiecon}. A possible remedy for this problem is including the negative pressure induced by loading cracks. 

The spatial dependence of the stress field at the crack tip is
\begin{align}
  p^{\rm (m)}(\Vec{r}) =-\frac{1}{3}\sigma_{kk}(\Vec{r})=-\frac{(2+\nu)}{3\sqrt{2\pi}}\frac{K_{\rm I}}{\sqrt{r}}\sin{\frac{\theta}{2}} 
\end{align}
where $r$ and $\theta$ are the polar coordinates of the point of interest with respect to the crack tip (cf.\ Figure~\ref{fig:atomistic}a1) and $\nu$ and $K_{\rm I}$ are the Poisson's ratio and the mode I stress intensity factor, respectively. The cross sectional area of the cylindrical domain around the crack tip at which the magnitude of the negative pressure is larger than the critical value $p^{\rm (m)}_{\rm crit}$ obtained from Figure ~\ref{fig:effic}a will be relevant for hydride formation. Figure ~\ref{fig:effic}b depicts the dependence of this area, $A$, on the external hydrogen gas pressure for different stress intensity factors of the crack tip at room temperature. We consider the hydrides with the cross sectional area of 10$\pi$ ${\rm \text{\AA}}^2$ as a critical size, as it ensures the presence of second nearest neighbor H atoms of an inner H atom in the hydride.  The determination of the critical nucleation size of the hydride requires careful calculation of the hydride/Al interface energy which is beyond the scope of this study.  The hydrides experimentally observed in this work, which form near the dislocations in crack tip, have a cross sectional area of $\approx 1000$ ${\rm \text{\AA}}^2$. 

The presence of a stress intensity factor significantly reduces the external H$_2$ gas pressure needed for overcoming this critical nucleus size. 
In the absence of the mechanical loading, i.e., $K_{\rm I}=0$, the gas pressure needed for the formation of hydrides is 349 MPa.  However, at this pressure the formation of the hydride is not local and thus the radius of the hydride grows unboundedly.  
%There is no local hydride formation in this case in the lower gas pressures.
 
The minimum value of the external pressure required for the formation of \emph{local} hydrides with $A=10\pi\text{\AA}^2$ reduces to $\approx 112$ MPa if a K-load of $\approx 0.05$ MPa$\sqrt{\rm m}$ is applied. By increasing the stress intensity factor to $\approx 0.15$ MPa$\sqrt{\rm m}$ this value decreases to $\approx 2$ MPa. For all larger K-loads, with application of 1 MPa of H$_2$ gas pressure we can get the minimum sized hydride. In cases at which the cracks are located along the ductile orientations the critical load for the emission of dislocations is $\approx 0.30$ MPa$\sqrt{\rm m}$~\cite{yamakov2014investigation}. 
Since the emission of the dislocations blunts the crack tip, this value is the maximum attainable K-load in the ductile direction. Under this load the critical H$_2$ gas pressure is reduced below 1 MPa. 
A further reduction of the required gas pressure is possible, since in realistic systems the maximum load can vary between $0.30$ MPa$\sqrt{\rm m}$ to loads less than $\approx 0.5$ MPa$\sqrt{\rm m}$, which is the cleavage load for the cracks along brittle orientations. 

The result shown in Figure~\ref{fig:effic}b indicates  that the negative pressure field of the edge dislocation has also an enhancing effect on the hydride formation~\cite{von2011hydrogen}. However, this effect cannot be tailored by mechanical loading and is not sufficient to form the experimentally observed hydrides.

\section{H accumulation in the crack tip region}\label{sec:crack}
The fundamental assumption behind the aforementioned analysis is that there exist enough H atoms in the vicinity of the defect to form the hydride phase. 
Using the example of cracks in Al, we now demonstrate the enhancing effect of the stress field next to microstructural defects on the energetics as well as the formation \emph{kinetics} of regions with high H concentration  at finite-temperature. 
Our MD--MC simulations are concerned with the accumulation of the H atoms at the crack tip region and the formation of any hydride nucleus, not necessarily  the Alane hydride. 
\subsection{Details of MD--MC simulations}

To demonstrate the above-described mechanism we modified the previously presented ADP potential of Al--H to include the 1NN and 2NN attractive interaction energies.  Based on  this potential,  Molecular Dynamics -- Monte Carlo (MD--MC) simulations are performed using the scalable parallel Monte Carlo algorithm~\cite{sadigh2012scalable}. The latter explores the phase space in a semi-grand canonical ensemble. The simulations are performed using the Large-scale Atomic/Molecular Massively Parallel Simulator (LAMMPS)~\cite{plimpton1995fast} and all atomistic Figures are created using the Open Visualization Tool~\cite{stukowski2009visualization}.  Each MC step tries to swap an empty T-site with a T-site occupied by a H atom using the Monte-Carlo energy criterion.

To examine the enhancement of hydride formation in the defected regions it is necessary to perform two sets of simulations.  In the first set, the samples consist of a defect-free $[l_x\times l_y\times l_z]\approx[316{\rm \text{\AA}} \times 463 {\rm \text{\AA}} \times 17.1 {\rm \text{\AA}}]$ box of bulk Al-containing
$\approx 153000$ Al atoms.  The crystallographic of $x$-,$y$- and $z$-directions are parallel to $[\bar{1}\bar{1}2]$, $[111]$, $[\bar{1}10]$, respectively.  The samples are equilibrated at 300 K for 50 ps.  Afterward, the MD--MC simulations are performed and the concentration of H atoms in the sample is recorded. In these simulations, using the variance-constrained semi-grand-canonical  ensemble~\cite{sadigh2012scalable}, the chemical potential is kept fixed.  The temperature is fixed at room temperature, $T=300$ K with the Nos\'e--Hoover thermostat~\cite{nose1984unified}. The time step for the integration in MD simulations is 0.2 fs.  Choice of this particularly small integration step ensures smooth behavior of light H atoms after addition to the system.  The  integration steps less than 0.5 fs were used in literature in similar studies~\cite{song2013atomic}.   The simulation time is 150 ns and $7.5\times 10^{6}$ MC steps are performed.

In the second set of simulations a $(111)[\bar{1}10]$ crack is inserted in the aforementioned sample by omitting the interactions between atoms at the opposite crack faces.% as shown in Figure~\ref{fig:geom}. 
The atoms of the sample are displaced according to the asymptotic continuum solution for a semi-infinite mode-I crack in an anisotropic linear elastic crystal~\cite{wu2015brittle}. The input parameters for this displacement field are the elastic constants of Al and the mode-I stress intensity factor $K_{\rm I}$. 

Using the ADMP potential the elastic constants of Al are computed as $C_{11}=114$~GPa, $C_{12}=61.6$~GPa and $C_{14}=31.6$~GPa.  
$k_I$ is kept less or equal to 0.30 MPa$\sqrt{\rm m}$, which is the critical load for the emission of  dislocations in pure Al.  
The boundaries of the sample are kept fixed and subsequently the samples are equilibrated at 300 K for 50 ps.  After the equilibriation, the above-described MD--MC simulations are performed.  The total simulation time is 5 ns and $2.5\times 10^{5}$ MC steps are performed in each simulation.  This number of steps was sufficient to observe the hydride embryos at the crack tip. 
After performing, the MD--MC simulation we relaxed the whole sample using the conjugate gradient method~\cite{hestenes1952} to attain the final structure. 

It should be noted that at the first step we excluded the free surfaces of the cracks from the domain of hydride embryo formation.  This modification enables us to only focus on the role of the stress field in facilitating hydride embryo formation.  This facilitating effect will be generalized to the other types of defects which are not necessarily associated with surfaces.  Moreover,  we are concerned with the micro-cracks usually along the grain boundaries inside the material, thus the notion of having a pure Al surface at the crack faces is not reasonable and the faces of the cracks will contain impurities.

\begin{figure}[ht]
   \centering
   \begin{subfigure}[b]{1\textwidth}
     \centering
     \includegraphics[width=1\textwidth]{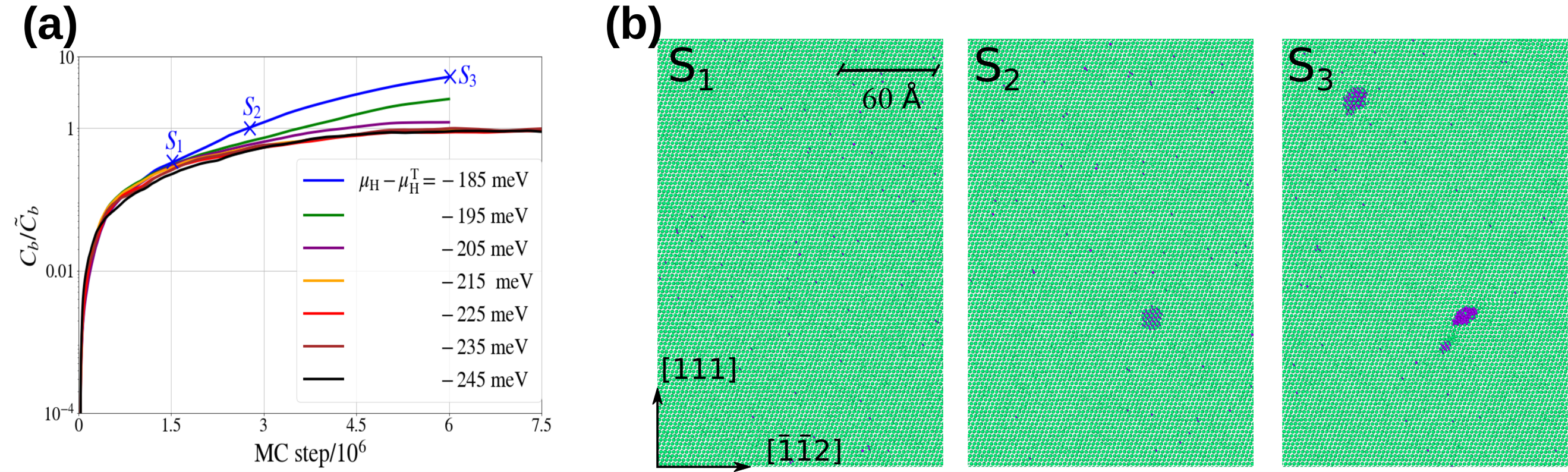}
   \end{subfigure}
   \centering

    \caption{(a) Evolution of the H concentration vs.\ MC steps in the MD--MC simulations for different chemical potentials of H at T=300K. The cross symbols indicate the MC steps at which (b) snapshots of atomistic configurations for a subset of a bulk sample with $\mu_{\rm H}-\mu^{\rm T}_{\rm H}= $-185 meV are taken and depicted in (b). The green and purple spheres denote the Al and H atoms, respectively. The snapshots show the nucleation and growth of hydride precipitates.}
    \label{fig:res:bulk}
\end{figure}

Figure \ref{fig:res:bulk}a shows the development of the overall atomic concentration, $C_{\rm b}$, of H atoms in Al bulk samples versus the MC attempts in the MD--MC simulations at room temperature. The chemical potential $\mu_{\rm H}-\mu^{\rm T}_{\rm H}$ is kept fixed using a scalable semi-grand canonical MC algorithm~\cite{sadigh2012scalable}. The concentrations are normalized to the theoretical value of bulk dilute non-interacting H atoms at a given chemical potential
\begin{align}
  \tilde{C}_b=\rho_s\exp\left((\mu_{\rm H}-\mu^{\rm T}_{\rm H})/k_{\rm B}T\right), 
\end{align}
where $\rho_s=2$ is the ratio of tetrahedral sites to lattice sites of fcc Al and $\mu_{\rm H}-\mu^{\rm T}_{\rm H}$ is the difference of the chemical potential and energy change due to insertion of H in the bulk Al, i.e., $\mu_{\rm H}-\mu^{\rm T}_{\rm H}=\mu_{\rm H}-E_{\rm sol}^{\rm T}-\mu_{\rm H}^0$, respectively. It can be seen that the concentration ratio converges to the constant value of 1 after $10^5$ MC steps for $\mu_{\rm H}-\mu^{\rm T}_{\rm H}\le -215$~meV. 

For higher chemical potentials,  the overall observed concentrations exceed the theoretical prediction as time progresses. This is an indication of a hydride  formation in addition to the H in solid solution. Figure~\ref{fig:res:bulk}b illustrates a subset of a crack-free sample with $\mu_{\rm H}-\mu^{\rm T}_{\rm H}=-185$ meV at three different MC steps. 
It can be seen that regions with high H concentrations form at random locations of the sample. 
These hydride \emph{nuclei} have the structure of fluorite AlH$_2$. The thermodynamic DFT considerations in Sec.~\ref{sec:thermo} suggest that they will relax to the AlH$_3$ hydride once the hydride becomes large enough to overcome the coherency strain.  
The thermodynamic prediction does not change in the pressure domain $p^{\rm(m)}\le 19.5$ GPa. The reason is that the volume per unit formula of AlH$_2$ is 3.585 $\text{\AA}^3$.  After the decomposition into 2/3AlH$_3$+1/3Al it changes to +8.183 $\text{\AA}^3$. As the corresponding gain in the energy  is -0.996 eV per Al atom, the lower bound pressure required for inhibiting this transformation by the $p^{\rm(m)}\Delta V$ term is 19.5 GPa, which is far beyond the attainable pressures. Performing the same analysis using the results of the ADMP gives rise to a critical pressure of $p^{\rm(m)}\le 4.5$ GPa.  To have a full picture, the large coherency strain of the AlH$_3$ hydride should be also taken into account.

 \begin{figure}
   \centering
  
     \includegraphics[width=0.8\textwidth]{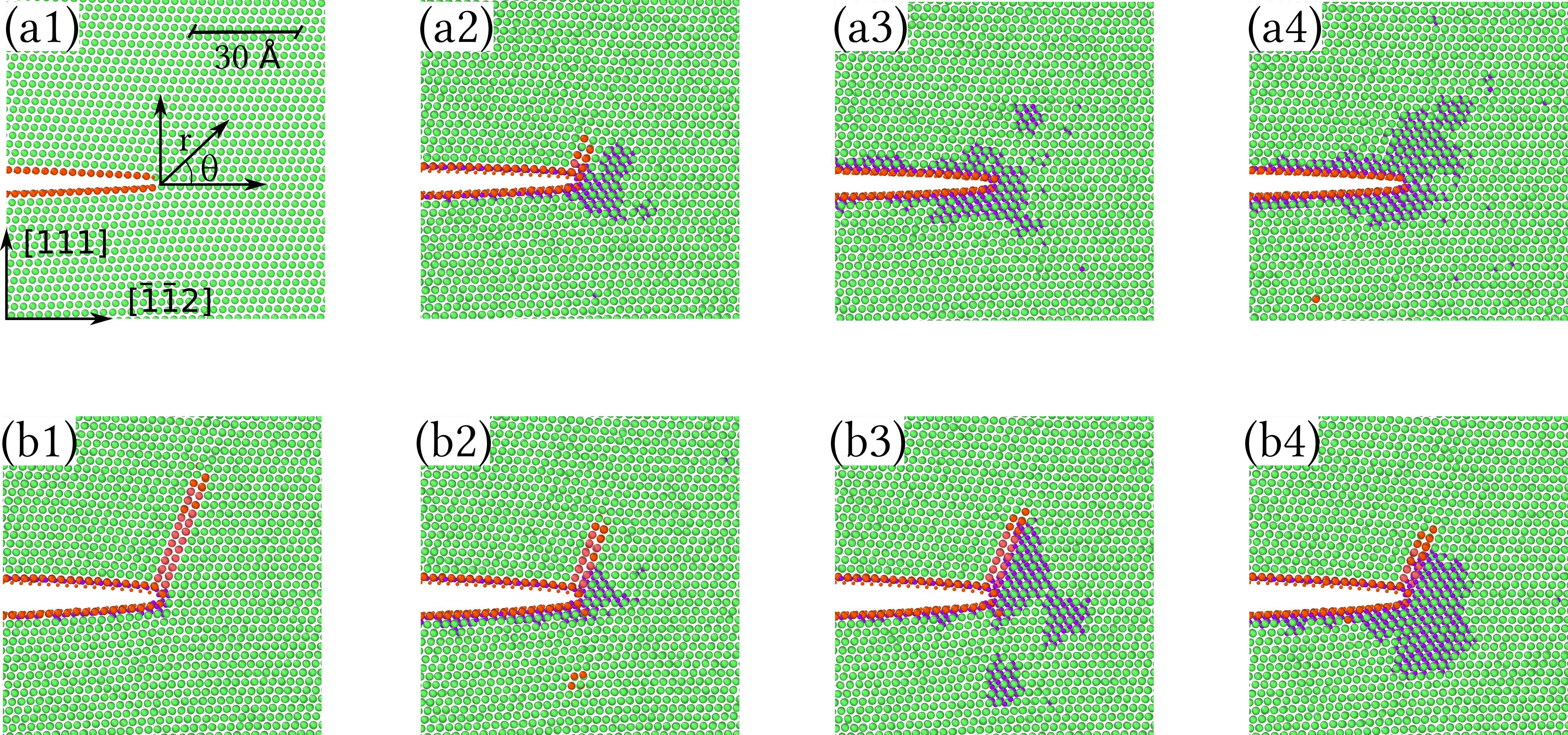}
    \caption{(a1)--(a4) Atomic structure of the crack tip regions at $T = 300$ K and $k_I=0.20$ MPa$\sqrt{\rm m}$ and $\mu_{\rm H}-\mu^{\rm T}_{\rm H} =$ -245 meV, -225 meV, -205 meV, -185~meV, respectively. (b1)--(b4) Same at $k_I = 0.30$ MPa$\sqrt{\rm m}$. The green, red and purple spheres denote the fcc-Al, non-fcc Al, and H atoms, respectively. }
    \label{fig:atomistic}
\end{figure}
To analyze the defected structures, Figure~\ref{fig:atomistic} illustrates the relaxed state of the hydride formed in the crack tip region at $T =300$ K after $2.5\times 10^{6}$ of MC steps.  It can be seen that the hydride forms at the crack tip at both $K_{\rm I}=0.20$ MPa$\sqrt{\rm m}$, 0.30 MPa$\sqrt{\rm m}$ and chemical potentials higher than $\mu_{\rm H}-\mu^{\rm T}_{\rm H}=-225$~meV. 
Since for this particular chemical potential, the bulk sample did not form a hydride nucleus after 7.5$\times 10^{6}$ MC steps, one can conclude that the formation of the hydride is enhanced due to the excess volume of H atoms and the stress field of the crack. 
Therefore, the enhancement of the hydride formation energetics at the crack tip due to its stress field is also accompanied by an increased concentration of the individual H atoms in this particular region and this accumulation enhances the kinetics of hydride formation. More precisely, in the case of the bulk samples the hydride nucleus at $\mu_{\rm H}-\mu^{\rm T}_{\rm H}=-0.185$ meV forms after 2.65$\times 10^6$ MC steps, while in the defected sample with $K_{\rm I}=0.3$ MPa$\sqrt{\rm m}$, this nucleus forms  at  $\mu_{\rm H}-\mu^{\rm T}_{\rm H}=-0.225$ meV and after 2.65$\times 10^6$ MC steps.

Interestingly, it can be seen that in all cases with $k_I=0.30$ MPa$\sqrt{\rm m}$, i.e., close to the previously observed emission loads at 300 K~\cite{yamakov2014investigation}, a partial dislocation is emitted.  Its emission in addition to the stacking fault didn't hinder the formation of a hydride nucleus. 

 \section{Experimental observation of H enriched regions near crack tips in a 7449-Al alloy }\label{sec:APT}
 \subsection{Technical details of atom probe tomography}
\begin{figure}[ht]\centering
 \includegraphics[width=\textwidth]{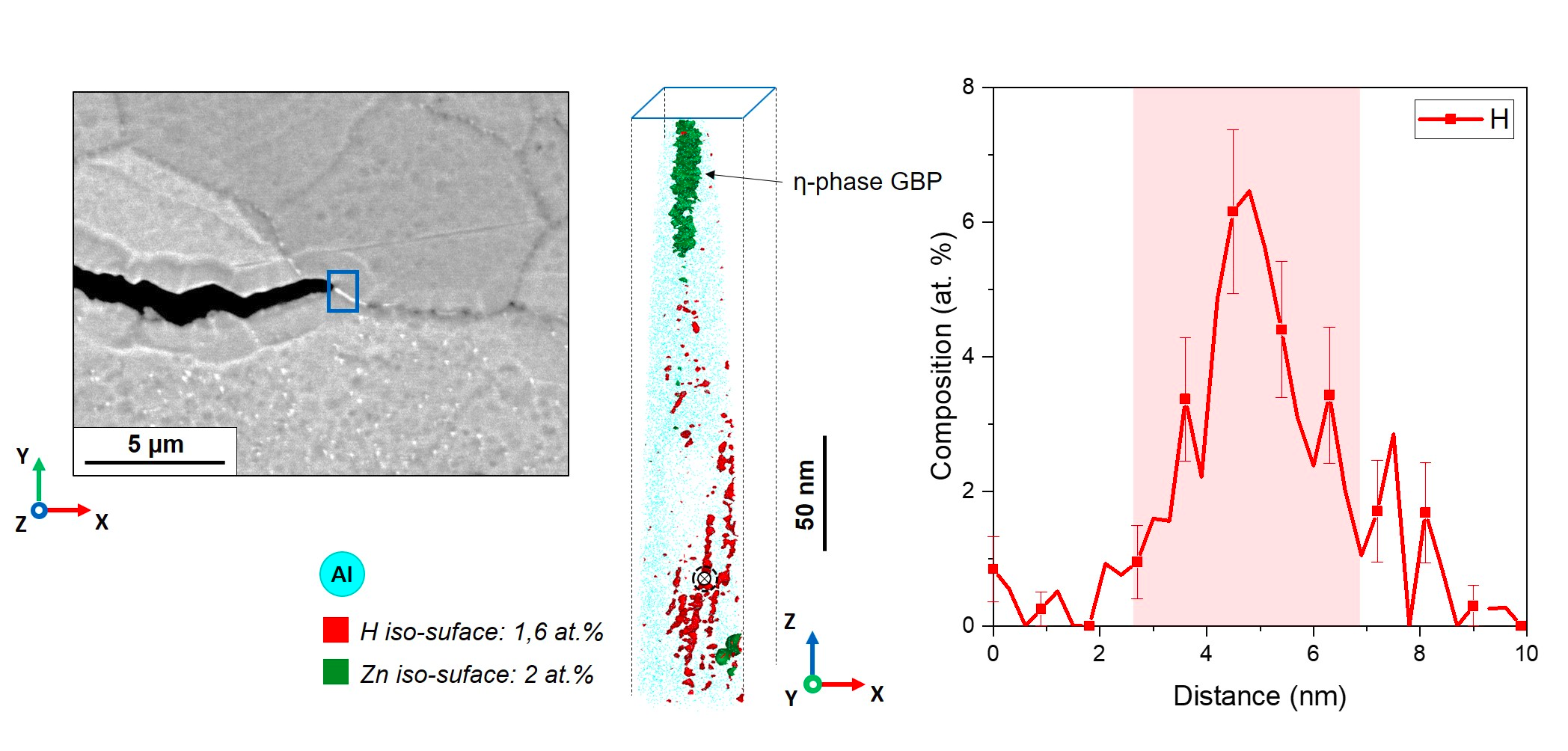}
    
    \caption{(left) SEM image depicting the tip of a main crack in a 7449-Al alloy. (middle) APT reconstruction near a stress-corrosion crack tip presenting elongated features on a (111) plane. (right) Composition profile along the y-axis at the position indicated by the cross in the APT tip. } \label{fig:APT}
\end{figure}

A stress-corrosion cracking, double cantilever beam (DCB) test was performed at room temperature on a commercial 7XXX-aluminium alloy (7449) following the ASTM G168-17 protocol~\cite{ASTM168}. Pre-cracking was done manually by “pop-in” and droplets of 3.5\% NaCl solution were added twice a day in the slit during weekdays. A series of specimens for atom probe tomography was prepared in the vicinity of the crack tip using a dual beam scanning-electron microscope - Xe-Plasma focused ion beam (FEI Helios PFIB) following the procedure outlined in Ref.~\cite{gault2018interfaces}. APT data was acquired by a reflectron fitted Local Electrode Atom Probe (LEAP 5000 XR, Cameca Instrument Inc. Madison, WI, USA). We used the voltage-pulsing mode, with a 20\% pulse fraction, a pulse repetition rate of 200 kHz. The specimen was maintained at a base temperature of 50K. The average detection rate was set at 35 ions detected per 10.000 pulses. The data pas processed and reconstructed using AP Suite 6.0, and the tomographic reconstruction was calibrated using the method introduced in Ref.~\cite{gault2009advances}. 
The scanning electron micrograph (SEM) in Figure~\ref{fig:APT} shows the tip of the main crack in the double cantilever beam sample, typically used to study stress-corrosion cracking of Al-alloys. The APT sample is prepared within $\approx$1 $\micro$m from the crack tip. The point cloud obtained from the reconstruction of the APT data is displayed, with blue points corresponding to Al atoms. 
The green isocomposition surfaces delineate Zn-rich regions ($>$ 2at.\%), indicating a large precipitate ($\eta$ phase) located at a grain boundary along which the crack is propagating. Regions rich in hydrogen ($>$1.6 at.\%) are highlighted by a set of red isosurfaces. They form a pattern located in the (111) planes as identified by atom probe crystallography approaches~\cite{gault2012atom, Zhou2021}. Measuring H by APT is notoriously challenging~\cite{meier2021extending,Sundell2013hydrogen,breen2020solute}, but we used experimental conditions that minimize the signal from the residual gas in the atom probe chamber. The strong, localized signal can be ascribed to the accumulation of H atoms in the crack tip region along parallel tubular features that were identified by correlative transmission-electron microscopy and APT ~\cite{Kuzmina2015linear,makineni2018diffusive, Zhou2021}. We measure H compositions, along a cylindrical region of interest. Here, the H ratio reaches over 6 at.\%.  Across the multiple datasets acquired, hydrogen ratios of up to 20 at.\% and more were recorded at these features,  contrasting  the concentrations ($<$1 at.\%) measured away from the crack. The distribution of H atoms can be attributed to segregation or to the formation of hydride nuclei at these linear defects.

These results provide experimental evidence that hydrogen tends to be in the stressed regions in the vicinity of the cracks. Here, hydrogen originates from the splitting of water on the freshly formed Al surface exposed by the progression of the crack, i.e., its ingress is likely of a different magnitude as compared to the cases studied by our atomistic approach. Nevertheless, the theoretical prediction of high equilibrium concentration of H atoms at the crack tip region is still valid. 

The observation of a microstructure that appears to be a set of parallel dislocations \cite{Zhou2021} contained within a (111) plane suggests that hydrogen has a tendency to cause this organisation: The presence of the crack and the induced negative pressure attracts H atoms.  These H atoms now find the pre-existing dislocations at the tip region, accumulate in their local tensile field  and form the hydride. Thus the formation of the high concentration regions  is a consequence of the combined effect of crack tip and dislocation stresses. 
At the same time, the presence of H reduces the distance of piled-up dislocations, by a shielding of the repulsive dislocation-dislocation interaction \cite{von2011hydrogen}.
Moreover, the formation of piled-up dislocations in the crack tip region is connected with stacking faults along the (111) planes, for which the DFT calculations depicted in Figure~\ref{fig:phase_for}b, and Figure~\ref{fig:planarEf} also predict an attraction for H atoms. Out of the planar hydrides, those next to stacking faults even have the lowest formation energy. 

The planar hydrides in the glide plane remain stable even after carving the APT needle out of the sample, which relaxes the crack tip stress, but not the tensile stress of dislocations. Since the formation energy of planar hydrides at (111) plane is low itself (cf. Figure~\ref{fig:planarEf}), the preference of H atoms to form hydrides along (111) planes as discussed in Sec.~\ref{sec:thermo} can be understood.  It should be noted that the hydrides are not observed at the dislocations far from the crack tip region. The reason is that the negative pressure of single dislocations is too low to thermodynamically stabilize  the formation of hydrides with the experimentally observed cross sectional area at hydrogen pressures lower than 20 MPa (cf. Figure~\ref{fig:effic}b). 

\section{Summary}\label{sec:discussion}
Using ab initio and atomistic calculations
and supported by experimental results, we have shown that an  enhanced formation of $\alpha$-Alane hydride at the negatively pressurized regions in Al is feasible. Using that example, we proved the concept of defects as laboratory tools for exploring the negative pressure domain of the phase diagram. In particular, the unique property of cracks to continuously and measurably change the negative pressure in the material when the applied loads are modified, provides an opportunity to systematically investigate the properties of hydrides. Cracks also show a unique potential for facilitating the charging and discharging process of metal hydrides. These novel investigations are crucial for tackling major technological problems arising against using hydrogen as a energy carrier. 
 
We herewith propose a route for tailoring negative pressure through defects, which can be considered as a remedy for enhancing metal hydride formation.  The information about the critical stress intensity factors for emission of dislocations and decohesion in Al can be used for the determination of a safe magnitude of external applied load by which the hydrides form prior to the propagation or blunting of cracks.  

The presence of suitable alloying element can ensure the presence of initial cracks with a suitable length in metal.  This enhancing effect, has certain implications for H-storage applications.  On the other hand, the interaction of hydrogen next to cracks and dislocations is also decisive for the study of hydrogen embrittlement.

For example, previous theoretical studies suggest that nano-hydrides can induce ductile-to-brittle transitions~\cite{song2011nanoscale,song2013atomic} in nickel. On the other hand, the presence of hydrides can facilitate the nucleation of dislocations~\cite{leyson2016multiscale} and probably make the material more locally ductile~\cite{birnbaum1994hydrogen}. 

Thus, using cracks as a tool for investigating the behavior of hydrides opens a new horizon for scientific and technologically relevant research.

 \section*{Acknowledgement}
 A.T, T.H, and J.N acknowledge financial support by the Deutsche Forschungsgemeinschaft (DFG) through the
projects A06 and C05 of the CRC1394 ``Structural and Chemical Atomic Complexity -- From Defect Phase Diagrams to Material Properties'', project ID 409476157.
P.C, M.L.F, and B.G  acknowledge financial support from the ERC-CoG-SHINE-771602.
 \appendix

\section{Solubility of H in Al}\label{App:A}
\begin{figure}[H]\captionsetup{type=table}
\caption{Selected properties of H in Al calculated by DFT, an angular dependent potential (ADP)~\cite{apostol2010angular} and an augmented ADP in conjunction with a Morse potential (ADMP).  $\Delta E_{\rm sol}^{T\rightarrow O}=E_{\rm sol}^{\rm O}-E_{\rm sol}^{\rm T}$ is the difference between the solution energy of H atoms in tetrahedral and octahedral sites. $E_{\rm int}^{i,j,mNN}$ is the interaction energy of sites $i$ and $j$  in the $m$th nearest neighbor configuration.  The numbers in brackets denote the formation energy of pairs w.r.t individual isolated H atoms in T-sites. $E_{\rm f}({\rm AlH}_m)$ is the formation energy per formula unit (f.u.) of the AlH$_m$ phase at $\Delta\mu=0$.  $\Delta V$ denotes the excess volume of H atoms, and the hydrides in the Al matrix. The different sites are depicted in the figure.}\label{Tab:int}
%\centering
\begin{minipage}{0.3\textwidth}
\centering
  \includegraphics[width=\textwidth]{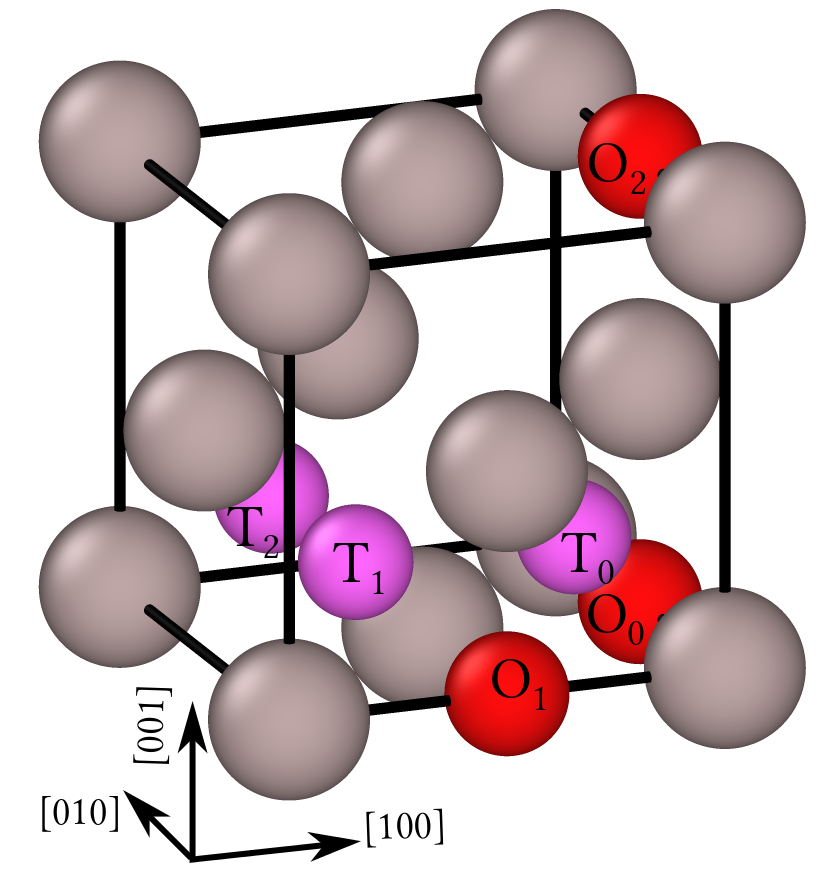}
  \captionsetup{type=figure}
\caption{Various tetrahedral (purple spheres) and octahedral (red) interstitial sites in Al (grey). }\label{fig:sites}
\end{minipage}
\begin{minipage}{0.5\textwidth}
\centering

\captionsetup{type=table} %% tell latex to change to table
 \begin{tabular}{cccc}
    \hline
        %\multicolumn{3}{c}{$E_{\rm int}^{i,j}$(eV)}\\
        quantity & DFT& ADP~\cite{apostol2010angular}&ADMP\\
         & (this work)  &(this work) &(this work)\\

        \hline
         $\Delta E_{\rm sol}^{T\rightarrow O}$(eV)&0.102&0.131&0.131\\
         $E_{\rm int}^{\rm T_0,T_1,1NN}$(eV)&-0.114&0.007&-0.114\\
         $E_{\rm int}^{\rm T_0,T_2,2NN}$(eV)&-0.028&-0.007&-0.031\\
         $E_{\rm int}^{\rm O_0,O_1,1NN}$(eV)&-0.111(0.093)&0.002&-0.057\\
         $E_{\rm int}^{\rm O_0,O_2,2NN}$(eV)&-0.061(0.143)&-0.016&-0.016\\
         $E_{\rm int}^{\rm O_0,T_0,1NN}$(eV)&-0.086(0.016)&0.031&-0.080\\
         $E_{\rm int}^{\rm O_0,T_1,2NN}$(eV)&-0.086(0.016)&-0.007&-0.027\\
         $E_{\rm f}({\rm AlH})$ (eV/f.u.)&0.740&0.784&0.637\\
         $E_{\rm f}({\rm AlH}_2)$ (eV/f.u.)&0.895&1.179&0.269\\
         $E_{\rm f}({\rm AlH}_3)$ (eV/f.u.)&-0.152&1.125&0.093\\
         $\Delta V$ (H) $(\text{\AA}^3)$&1.808&2.720&2.720\\
         $\Delta V$ (AlH) $(\text{\AA}^3)$&-0.027&1.763&-0.149\\
         $\Delta V$ (AlH$_2$) $(\text{\AA}^3)$&3.585&2.861&1.646\\
         $\Delta V$ (AlH$_3$) $(\text{\AA}^3)$&17.652&15.657&11.852\\              \hline
    \end{tabular}
\end{minipage}
\end{figure}

The solution enthalpy of single H atoms in Al as well as the interaction energies between various sites in Al calculated using DFT and empirical potential simulations  are given in Table~\ref{Tab:int}.  Moreover, the excess volume of the hydrides of interest are given in Table~\ref{Tab:int}.
\section{Planar Hydrides in fcc Al}\label{App:B}
\begin{figure}[H]
     \centering
         \includegraphics[width=0.8\textwidth]{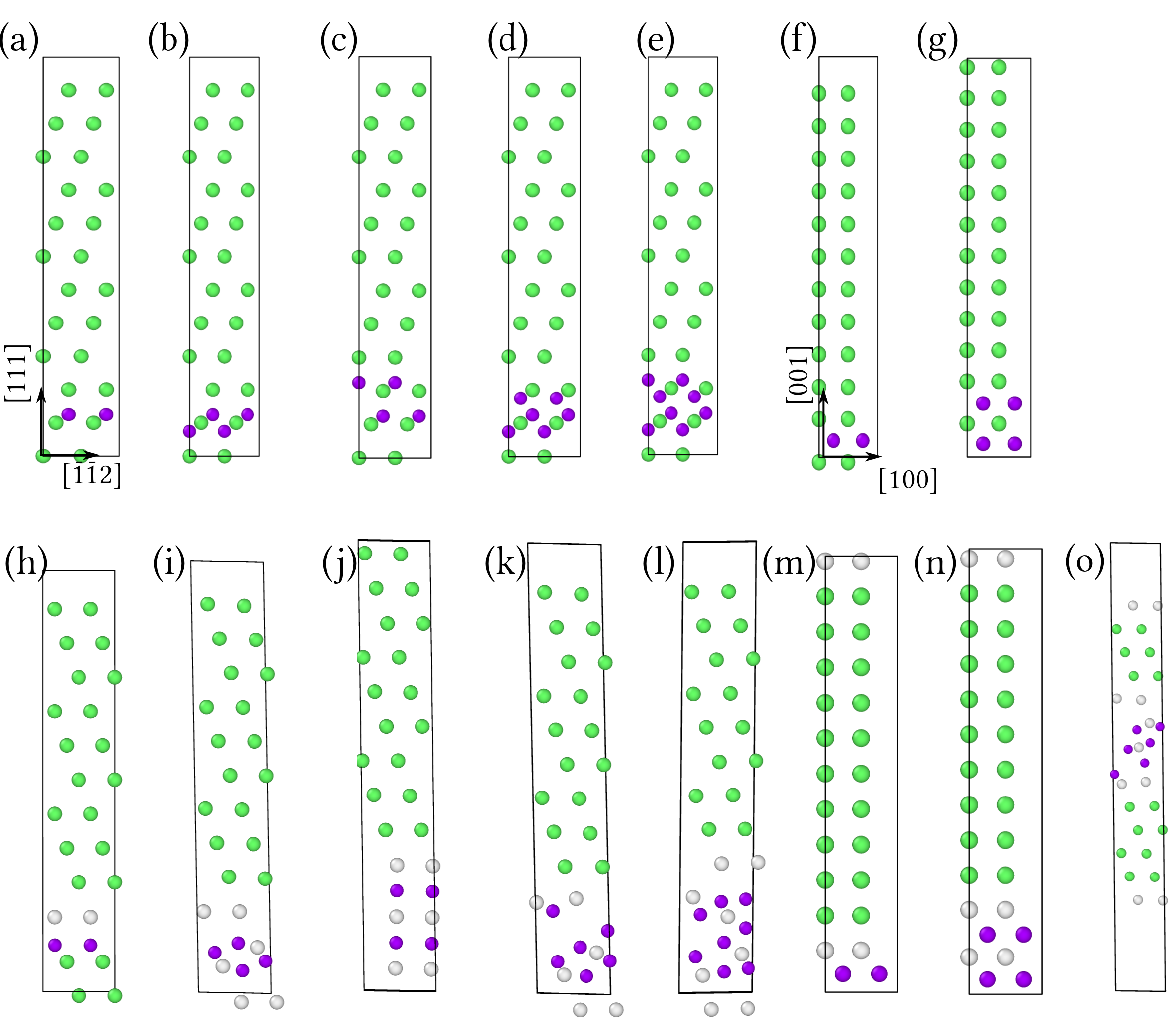}
        \caption{Initial structure of the supercell containing (a) 1 layer H along (111) , (b) 2 neighboring layers of H along (111), (c) 2 layers of  H along (111), (d) 3 neighboring layers of  H along (111), (e) 4 neighboring layers of  H along (111), (f) 1 layer of  H along (100), and (g) 2 neighboring layers of  H along (100) planes.  (h)--(n) denote the final structures of the aforementioned planar hydrides with the same order. (o) The relaxed structure of the supercell containing a stacking fault in a 3-layered hydride at (111) plane. The green, white, and purple spheres denote the fcc-Al, unrecognized-Al, and H atoms, respectively.  }
        \label{fig:planar}
\end{figure}

Several planar hydrides along (111) and (100) planes are created by filling the T-sites along these planes.  The initial and final structure of the planar hydride of interest are illustrated in Figure~\ref{fig:planar}.  

\begin{figure}[ht]
     \centering
    
         \includegraphics[width=\textwidth]{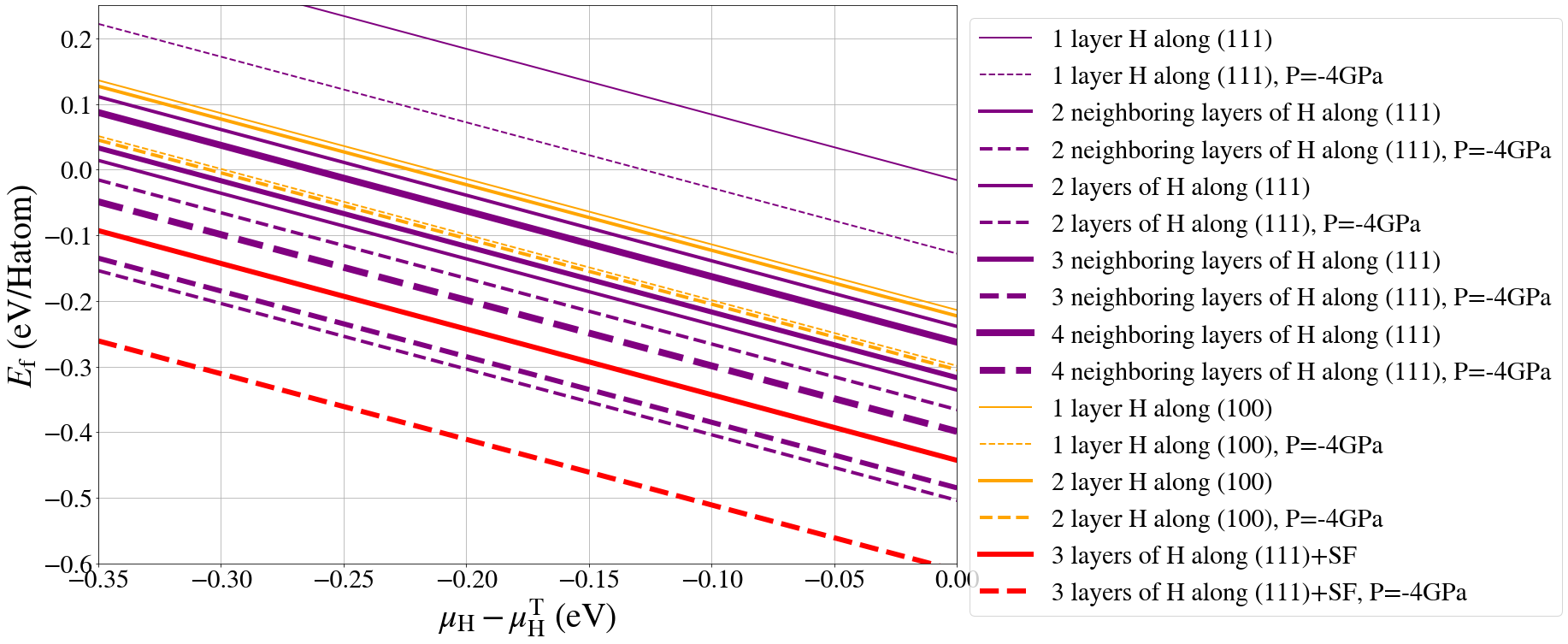}
        \caption{Formation energies of several planar hydrides per H atoms. The solid and dashed lines denote the formation energy in absence and presence of a negative pressure field. The planar structures are named using the number of the H layers and the plane along which the hydride is spread. The only exceptions are (111)2La and (1111)2Lb, which denote the 2-layered hydride along (111) plane as shown in Figure~\ref{fig:planar}b, and c, respectively.}
        \label{fig:planarEf}
\end{figure}

The formation energies per hydrogen atom of all of these structures in the absence and presence of the negative pressure field are given in Figure~\ref{fig:planarEf}.  It is evident that the 3 layers of H along (111)+SF defect which corresponds to a stacking fault surrounded by a three layered hydride is the most stable one. This hydride is comprised of H atoms in three successive (111) planes the initial and relaxed structure of this hydride without stacking fault are depicted in Figure~\ref{fig:planar}d and  Figure~\ref{fig:planar}k, respectively. The geometry of the relaxed supercell containing this three layered hydride and a stacking fault is given in Figure~\ref{fig:planar}o.  The stacking fault decreases the energy of the supercell by -0.126 eV/H atom.  These values are added to the formation energy of the hydride.
\section{The augmented angular dependent potential}\label{App:C}

A Morse type formulation is used to augment the H-H interaction in the ADP potential,
\begin{align}
   E_{\rm H-H}=D_0\left({\rm e}^{-2\alpha(r-r_0)} - 2{\rm e}^{-\alpha(r-r_0)}\right) , \label{Eq:morse}
\end{align}
where $D_0=0.134$ eV, $\alpha=2.42~\text{\AA}^{-1}$ and $r_0=2.02$~\AA.  The properties of pure Al and of a single H atom in the Al matrix calculated using this angular dependence in conjunction with the Morse potential (ADMP) remain unaltered with respect to the original ADP.  The results of this modification are also tabulated in Table~\ref{Tab:int} of the paper.  It can be seen that now the 1NN and 2NN H--H interaction energies are in agreement with the DFT results.  
\section{MD--MC simulation including the free surface of crack faces}\label{App:D}
\begin{figure}[ht]
     \centering
    
         \includegraphics[width=0.8\textwidth]{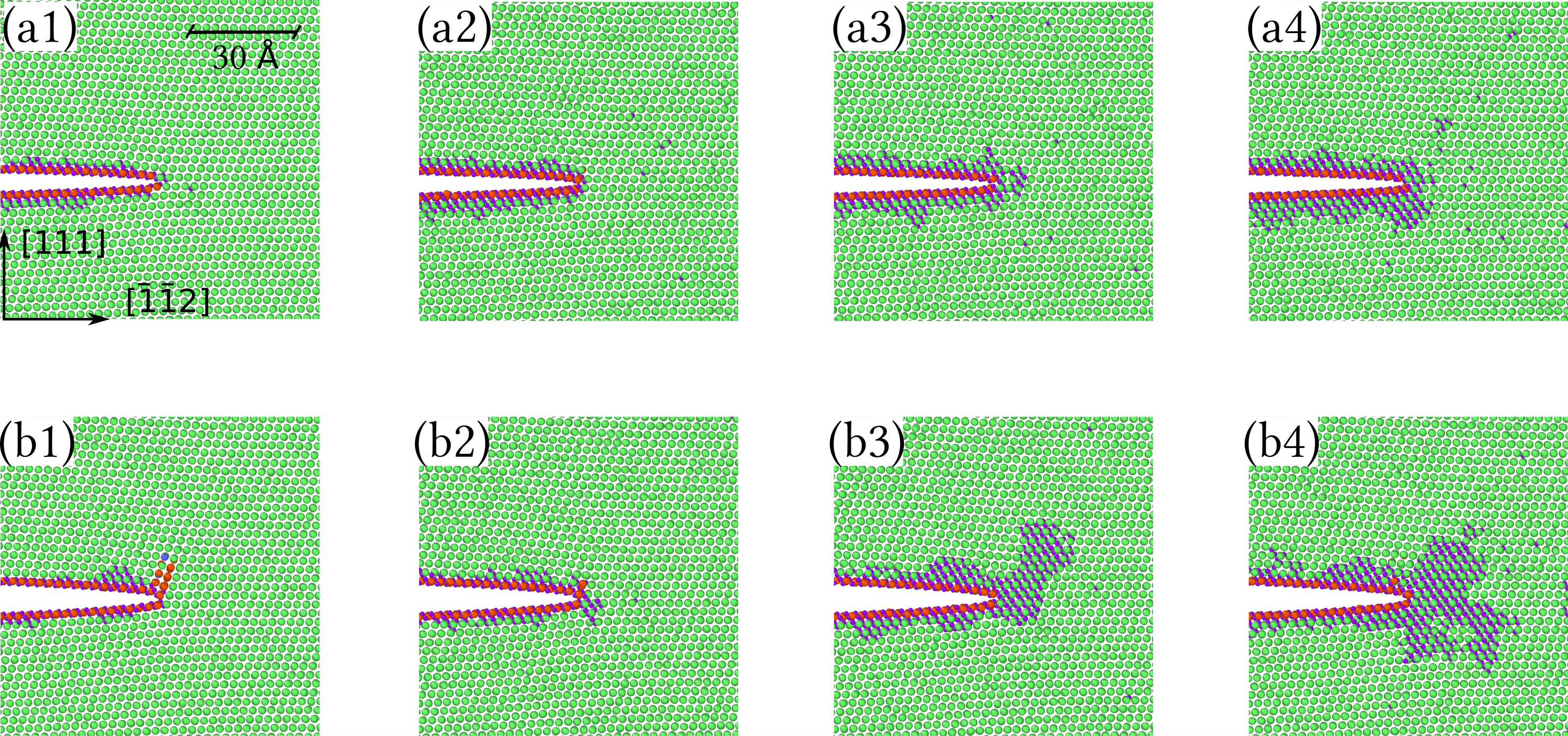}
        \caption{Same as Figure~\ref{fig:atomistic} of the paper, but with the surface sites included in the simulations.}
        \label{fig:atomistic_surf}
\end{figure}

For sake of comparison,  we add another simulation in which we don't exclude the free surfaces from the simulations. Figure~\ref{fig:atomistic_surf} depicts the key atomistic configuration at the crack tip simulations in this case.  It can be seen that the free surfaces are acting as formation nuclei for the hydrides. In this case we are observing the formation of hydrides at every k-load. In the cases with $\Delta \mu > -205$ meV, we observe that hydrides form at high stressed regions and along the crack faces. This observation demonstrates the enhancing effect of the free surfaces.  
 
\section{Vibrational free energy of \texorpdfstring{$\alpha$-Alane hydride}{Lg}}\label{App:E}
\begin{figure}[H]
    \centering
    \includegraphics[width=0.6\textwidth]{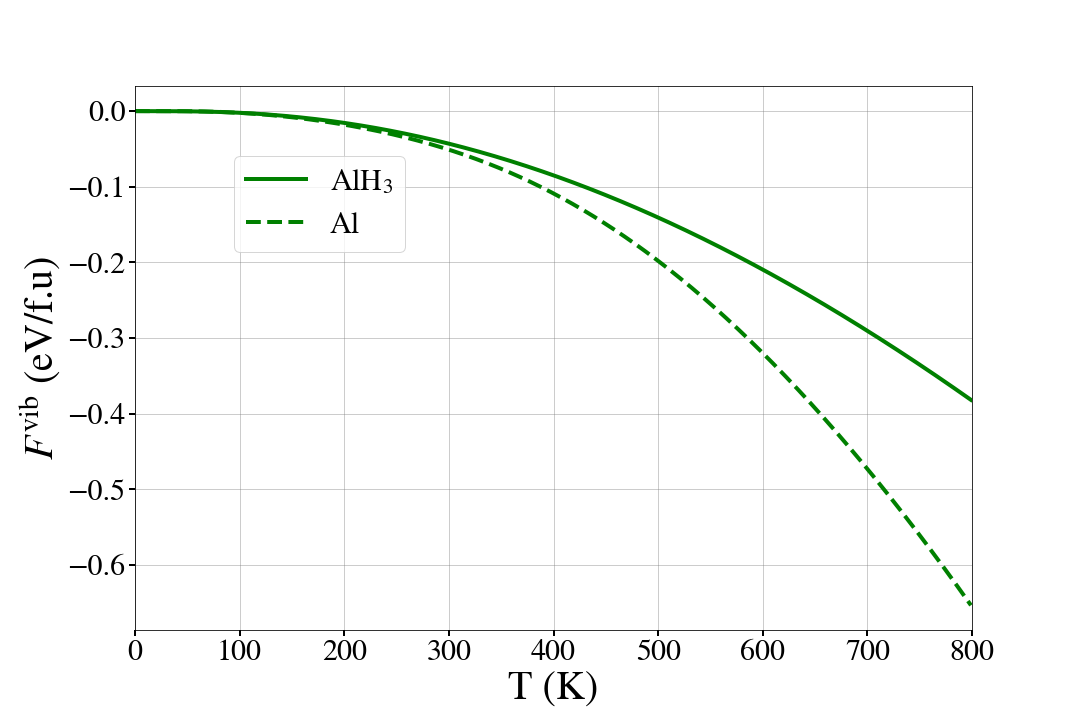}
    \caption{The vibration part of the formation free energies versus temperature.}
    \label{fig:vib}
\end{figure}

We calculated the vibrational free energy of the $\alpha$-Alane hydride and pure Al using the quasi-harmonic approximation~\cite{grabowski2007ab}.  The dynamical matrix components are calculated using DFT simulations of a Al$_6$H$_{18}$ and Al$_4$ supercells. Figure~\ref{fig:vib} shows the variation of the vibrational free energy versus temperature.  
The results of the APT experiments support the findings presented in the this set of  MD--MC simulations in which the enhancement of the H concentration due to the stress field of the crack is predicted.  The fact that in the experiments sources of H atoms are observed in crack faces doesn't contradict with the simulations since the crack surface and the stressed region in the bulk are in chemical equilibrium at a certain chemical potential.

\bibliography{bibliography.bib}

\begin{thebibliography}{10}
\expandafter\ifx\csname url\endcsname\relax
  \def\url#1{\texttt{#1}}\fi
\expandafter\ifx\csname urlprefix\endcsname\relax\def\urlprefix{URL }\fi
\expandafter\ifx\csname href\endcsname\relax
  \def\href#1#2{#2} \def\path#1{#1}\fi

\bibitem{smil2018energy}
V.~Smil, Energy and civilization: a history, MIT Press, 2018.

\bibitem{olivier2012trends}
J.~G. Olivier, G.~Janssens-Maenhout, J.~Peters, Trends in global co2 emissions,
  PBL Netherlands Environmental Assessment Agency 500114022 (2012) 6--39.

\bibitem{dawood2020hydrogen}
F.~Dawood, M.~Anda, G.~Shafiullah, Hydrogen production for energy: An overview,
  International Journal of Hydrogen Energy 45~(7) (2020) 3847--3869.

\bibitem{klebanoff2012final}
L.~Klebanoff, J.~Keller, Final report for the doe metal hydride center of
  excellence, Sandia National Laboratories, Albuquerque, NM, Report No.
  SAND2012-0786.

\bibitem{smith2021isotherm}
D.~B. Smith, R.~C. Bowman, L.~M. Anovitz, C.~Corgnale, M.~Sulic, Isotherm
  measurements of high-pressure metal hydrides for hydrogen compressors,
  Journal of Physics: Energy 3~(3) (2021) 034004.

\bibitem{imre2007existence}
A.~R. Imre, On the existence of negative pressure states, physica status solidi
  (b) 244~(3) (2007) 893--899.

\bibitem{mcmillan2003new}
P.~McMillan, New materials from high pressure experiments: challenges and
  opportunities, High Pressure Research 23~(1-2) (2003) 7--22.

\bibitem{mcmillan2002new}
P.~F. McMillan, New materials from high-pressure experiments, Nature materials
  1~(1) (2002) 19--25.

\bibitem{siol2018negative}
S.~Siol, A.~Holder, J.~Steffes, L.~T. Schelhas, K.~H. Stone, L.~Garten, J.~D.
  Perkins, P.~A. Parilla, M.~F. Toney, B.~D. Huey, et~al., Negative-pressure
  polymorphs made by heterostructural alloying, Science advances 4~(4) (2018)
  eaaq1442.

\bibitem{gupta2021combined}
A.~Gupta, B.~Tas, D.~Korbmacher, B.~Dutta, Y.~Neitzel, B.~Grabowski, T.~Hickel,
  V.~Esin, S.~V. Divinski, G.~Wilde, et~al., A combined experimental and
  first-principles based assessment of finite-temperature thermodynamic
  properties of intermetallic al3sc, Materials 14~(8) (2021) 1837.

\bibitem{jiang2021alh3}
W.~Jiang, H.~Wang, M.~Zhu, Alh3 as a hydrogen storage material: recent
  advances, prospects and challenges, Rare Metals 40~(12) (2021) 3337--3356.

\bibitem{graetz2013recent}
J.~Graetz, B.~C. Hauback, Recent developments in aluminum-based hydrides for
  hydrogen storage, MRS bulletin 38~(6) (2013) 473--479.

\bibitem{graetz2011aluminum}
J.~Graetz, J.~Reilly, V.~Yartys, J.~Maehlen, B.~Bulychev, V.~Antonov,
  B.~Tarasov, I.~Gabis, Aluminum hydride as a hydrogen and energy storage
  material: past, present and future, Journal of Alloys and Compounds 509
  (2011) S517--S528.

\bibitem{kresse1993ab}
G.~Kresse, J.~Hafner, Ab initio molecular dynamics for liquid metals, Physical
  Review B 47~(1) (1993) 558.

\bibitem{kresse1996efficiency}
G.~Kresse, J.~Furthm{\"u}ller, Efficiency of ab-initio total energy
  calculations for metals and semiconductors using a plane-wave basis set,
  Computational materials science 6~(1) (1996) 15--50.

\bibitem{kresse1996efficient}
G.~Kresse, J.~Furthm{\"u}ller, Efficient iterative schemes for ab initio
  total-energy calculations using a plane-wave basis set, Physical Review B
  54~(16) (1996) 11169.

\bibitem{perdew1996generalized}
J.~P. Perdew, K.~Burke, M.~Ernzerhof, Generalized gradient approximation made
  simple, Physical review letters 77~(18) (1996) 3865.

\bibitem{Methfessel1989}
M.~Methfessel, A.~T. Paxton, High-precision sampling for brillouin-zone
  integration in metals, Physical Review B 40~(6) (1989) 3616--3621.

\bibitem{wolverton2004hydrogen}
C.~Wolverton, V.~Ozoli{\c{n}}{\v{s}}, M.~Asta, Hydrogen in aluminum:
  First-principles calculations of structure and thermodynamics, Physical
  Review B 69~(14) (2004) 144109.

\bibitem{von2011hydrogen}
J.~von Pezold, L.~Lymperakis, J.~Neugebeauer, Hydrogen-enhanced local
  plasticity at dilute bulk h concentrations: The role of {H}--{H} interactions
  and the formation of local hydrides, Acta Materialia 59~(8) (2011)
  2969--2980.

\bibitem{birnbaum1997hydrogen}
H.~Birnbaum, C.~Buckley, F.~Zeides, E.~Sirois, P.~Rozenak, S.~Spooner, J.~Lin,
  Hydrogen in aluminum, Journal of Alloys and Compounds 253 (1997) 260--264.

\bibitem{apostol2010angular}
F.~Apostol, Y.~Mishin, Angular-dependent interatomic potential for the
  aluminum-hydrogen system, Physical Review B 82~(14) (2010) 144115.

\bibitem{tehranchi2017softening}
A.~Tehranchi, B.~Yin, W.~Curtin, Softening and hardening of yield stress by
  hydrogen--solute interactions, Philosophical Magazine 97~(6) (2017) 400--418.

\bibitem{yamakov2006molecular}
V.~Yamakov, E.~Saether, D.~R. Phillips, E.~H. Glaessgen, Molecular-dynamics
  simulation-based cohesive zone representation of intergranular fracture
  processes in aluminum, Journal of the Mechanics and Physics of Solids 54~(9)
  (2006) 1899--1928.

\bibitem{van2004thermodynamics}
A.~Van~der Ven, G.~Ceder, The thermodynamics of decohesion, Acta Materialia
  52~(5) (2004) 1223--1235.

\bibitem{grabowski2007ab}
B.~Grabowski, T.~Hickel, J.~Neugebauer, Ab initio study of the thermodynamic
  properties of nonmagnetic elementary fcc metals: Exchange-correlation-related
  error bars and chemical trends, Physical Review B 76~(2) (2007) 024309.

\bibitem{atkins1998physical}
P.~W. Atkins, J.~De~Paula, Physical chemistry, Oxford university press, Oxford
  UK, 1998.

\bibitem{san2007permeability}
C.~San~Marchi, B.~P. Somerday, S.~L. Robinson, Permeability, solubility and
  diffusivity of hydrogen isotopes in stainless steels at high gas pressures,
  International Journal of Hydrogen Energy 32~(1) (2007) 100--116.

\bibitem{yamakov2014investigation}
V.~Yamakov, D.~Warner, R.~Zamora, E.~Saether, W.~Curtin, E.~Glaessgen,
  Investigation of crack tip dislocation emission in aluminum using multiscale
  molecular dynamics simulation and continuum modeling, Journal of the
  Mechanics and Physics of Solids 65 (2014) 35--53.

\bibitem{ciaraldi1980studies}
S.~Ciaraldi, J.~Nelson, R.~Yeske, E.~Pugh, Studies of hydrogen embrittlement
  and stress-corrosion cracking in an aluminum-zinc-magnesium alloy.[5. 6 zn-2.
  6 mg], Tech. rep., Illinois Univ., Urbana (USA). Dept. of Metallurgy and
  Mining Engineering (1980).

\bibitem{scully2012hydrogen}
J.~Scully, G.~Young~Jr, S.~Smith, Hydrogen embrittlement of aluminum and
  aluminum-based alloys, in: Gaseous hydrogen embrittlement of materials in
  energy technologies, Elsevier, 2012, pp. 707--768.

\bibitem{zhang2021first}
D.-l. Zhang, W.~Jiong, K.~Yi, Z.~You, D.~Yong, First-principles investigation
  on stability and electronic structure of sc-doped $\theta$'/al interface in
  al- cu alloys, Transactions of Nonferrous Metals Society of China 31~(11)
  (2021) 3342--3355.

\bibitem{tsukahara1999improvement}
M.~Tsukahara, K.~Takahashi, A.~Isomura, T.~Sakai, Improvement of the cycle
  stability of vanadium-based alloy for nickel-metal hydride (ni--mh) battery,
  Journal of alloys and compounds 287~(1-2) (1999) 215--220.

\bibitem{tsukahara1997vanadium}
M.~Tsukahara, K.~Takahashi, T.~Mishima, A.~Isomura, T.~Sakai, Vanadium-based
  solid solution alloys with three-dimensional network structure for high
  capacity metal hydride electrodes, Journal of alloys and compounds 253 (1997)
  583--586.

\bibitem{wikiecon}
W.~contributors, \href{https://en.wikipedia.org/wiki/Hydrogen_economy}{Hydrogen
  economy}.
\newline\urlprefix\url{https://en.wikipedia.org/wiki/Hydrogen_economy}

\bibitem{sadigh2012scalable}
B.~Sadigh, P.~Erhart, A.~Stukowski, A.~Caro, E.~Martinez, L.~Zepeda-Ruiz,
  Scalable parallel monte carlo algorithm for atomistic simulations of
  precipitation in alloys, Physical Review B 85~(18) (2012) 184203.

\bibitem{plimpton1995fast}
S.~Plimpton, Fast parallel algorithms for short-range molecular dynamics,
  Journal of computational physics 117~(1) (1995) 1--19.

\bibitem{stukowski2009visualization}
A.~e. Stukowski, Visualization and analysis of atomistic simulation data with
  ovito--the open visualization tool, Modelling and {S}imulation in Materials
  Science and Engineering 18~(1) (2009) 015012.

\bibitem{nose1984unified}
S.~Nos{\'e}, A unified formulation of the constant temperature molecular
  dynamics methods, The Journal of chemical physics 81~(1) (1984) 511--519.

\bibitem{song2013atomic}
J.~Song, W.~Curtin, Atomic mechanism and prediction of hydrogen embrittlement
  in iron, Nature materials 12~(2) (2013) 145--151.

\bibitem{wu2015brittle}
Z.~Wu, W.~A. Curtin, Brittle and ductile crack-tip behavior in magnesium, Acta
  Materialia 88 (2015) 1--12.

\bibitem{hestenes1952}
M.~Hestenes, E.~Stiefel, Methods of conjugate gradients for solving linear
  systems, Journal of Research of the National Bureau of Standards 49~(6)
  (1952) 409.

\bibitem{ASTM168}
Astm g168-17: Standard practice for making and using precracked double beam
  stress corrosion specimens.

\bibitem{gault2018interfaces}
B.~Gault, A.~J. Breen, Y.~Chang, J.~He, E.~A. J{\"a}gle, P.~Kontis,
  P.~K{\"u}rnsteiner, A.~K. da~Silva, S.~K. Makineni, I.~Mouton, et~al.,
  Interfaces and defect composition at the near-atomic scale through atom probe
  tomography investigations, Journal of Materials Research 33~(23) (2018)
  4018--4030.

\bibitem{gault2009advances}
B.~Gault, M.~P. Moody, F.~De~Geuser, G.~Tsafnat, A.~La~Fontaine, L.~T.
  Stephenson, D.~Haley, S.~P. Ringer, Advances in the calibration of atom probe
  tomographic reconstruction, Journal of Applied Physics 105~(3) (2009) 034913.

\bibitem{gault2012atom}
B.~Gault, M.~P. Moody, J.~M. Cairney, S.~P. Ringer, Atom probe crystallography,
  Materials Today 15~(9) (2012) 378--386.

\bibitem{Zhou2021}
X.~Zhou, J.~R. Mianroodi, A.~{Kwiatkowski da Silva}, T.~Koenig, G.~B. Thompson,
  P.~Shanthraj, D.~Ponge, B.~Gault, B.~Svendsen, D.~Raabe,
  \href{https://advances.sciencemag.org/lookup/doi/10.1126/sciadv.abf0563}{{The
  hidden structure dependence of the chemical life of dislocations}}, Science
  Advances 7~(16) (2021) eabf0563.
\newblock \href {http://dx.doi.org/10.1126/sciadv.abf0563}
  {\path{doi:10.1126/sciadv.abf0563}}.
\newline\urlprefix\url{https://advances.sciencemag.org/lookup/doi/10.1126/sciadv.abf0563}

\bibitem{meier2021extending}
M.~S. Meier, M.~E. Jones, P.~J. Felfer, M.~P. Moody, D.~Haley, Extending
  estimating hydrogen content in atom probe tomography experiments where h2
  molecule formation occurs, Microscopy and Microanalysis (2021) 1--14.

\bibitem{Sundell2013hydrogen}
G.~Sundell, M.~Thuvander, H.-O. Andr{\'e}n, Hydrogen analysis in apt: methods
  to control adsorption and dissociation of h2, Ultramicroscopy 132 (2013)
  285--289.

\bibitem{breen2020solute}
A.~J. Breen, L.~T. Stephenson, B.~Sun, Y.~Li, O.~Kasian, D.~Raabe, M.~Herbig,
  B.~Gault, Solute hydrogen and deuterium observed at the near atomic scale in
  high-strength steel, Acta Materialia 188 (2020) 108--120.

\bibitem{Kuzmina2015linear}
M.~Kuzmina, M.~Herbig, D.~Ponge, S.~Sandl{\"o}bes, D.~Raabe, Linear
  complexions: Confined chemical and structural states at dislocations, Science
  349~(6252) (2015) 1080--1083.

\bibitem{makineni2018diffusive}
S.~K. Makineni, A.~Kumar, M.~Lenz, P.~Kontis, T.~Meiners, C.~Zenk,
  S.~Zaefferer, G.~Eggeler, S.~Neumeier, E.~Spiecker, et~al., On the diffusive
  phase transformation mechanism assisted by extended dislocations during creep
  of a single crystal coni-based superalloy, Acta Materialia 155 (2018)
  362--371.

\bibitem{song2011nanoscale}
J.~Song, W.~A. Curtin, A nanoscale mechanism of hydrogen embrittlement in
  metals, Acta Materialia 59~(4) (2011) 1557--1569.

\bibitem{leyson2016multiscale}
G.~Leyson, B.~Grabowski, J.~Neugebauer, Multiscale modeling of hydrogen
  enhanced homogeneous dislocation nucleation, Acta Materialia 107 (2016)
  144--151.

\bibitem{birnbaum1994hydrogen}
H.~K. Birnbaum, P.~Sofronis, Hydrogen-enhanced localized plasticity—a
  mechanism for hydrogen-related fracture, Materials Science and Engineering: A
  176~(1) (1994) 191--202.

\end{thebibliography}

\end{document}